\documentstyle[12pt, amsfonts]{article}
\begin{document}

%% typing started on 30.09.2002/Sofia
%% finished 17.11.2002/DFT-Univ. Trieste
%% corrections - till 27.11.2002 ESI Vienna
%% \hfill hep-th/0211154 (v.2 sent on 28.11.02)
%% ESI 1242
%% revised version for JPA - 31.01.2003/Sofia
%% revised version for JPA - 03.02.2003/Sofia

%%%%%%%%%%%%%%%%%%%%% (AMS)LATEX MACROS %%%%%%%%%%%%%%%%

\def\theequation{\thesection.\arabic{equation}}
\def\be{\begin{equation}}
\def\ee{\end{equation}}
\def\ba{\begin{eqnarray}}
\def\ea{\end{eqnarray}}
\def\lb{\label}
\def\nn{\nonumber}

\def\a{\alpha}
\def\b{\beta}
\def\g{\gamma}
\def\d{\delta}
\def\e{\varepsilon}
\def\l{\lambda}
\def\r{\rho}
\def\s{\sigma}
\def\o{\omega}
\def\O{\Omega}
\def\L{\Lambda}
\def\E{{\cal E}}
\def\Vp{{\cal V}_p}
\def\Hp{{\cal H}_p}
\def\p{\partial}
\def\bq{\overline{q}}
\def\bd{\bf d}
\def\bJ{{\bf J}}
\def\sbJ{^*\!{{\bf J}}}
\def\ux{\underline x}
\def\id{\mbox{\rm 1\hspace{-3.5pt}I}}
\newcommand{\ID}[2]{\id^{| #1 {\cal i}}_{\;\;\; {\cal h} #2 |}}

\def\C{\Bbb C}
\def\Z{\Bbb Z}
\def\F{\Bbb F}
\def\1{1\!\!{\rm I}}
\def\eod{\phantom{a}\hfill \rule{2.5mm}{2.5mm}}

\def\hF{\hat{F}}
\def\hA{\hat{A}}
\def\hB{\hat{B}}
\newcommand{\dy}[1]{DYBE(${#1}$)}
\newcommand{\xR}[1]{\ ^{\small #1}\!\!\hR}
\newcommand{\xF}[1]{\ ^{\small #1}\!\!\hF}

\newcommand{\rank}{\mathop{\rm rank}\nolimits}
\newcommand{\height}{\mathop{\rm height}\nolimits}
\newcommand{\aut}{\mathop{\rm Aut}\nolimits}
\newcommand{\rx}{\mathop{\rho_{\hspace{-1pt}\scriptscriptstyle W,k}}\nolimits}
\newcommand{\rn}{\mathop{\rho_{\hspace{-1pt}\scriptscriptstyle W,n}}\nolimits}
\newcommand{\q}[1]{[#1]}
\newcommand{\ai}[1]{{a_{#1}}\, }
\newcommand{\ainv}[2]{({a^{-1})}^{| #1 {\cal i}}_{\;\;\; {\cal h} #2 |}}

\def\R{\Bbb R}

\def\Rp{\hat{R}(p)}
\newcommand{\DR}[1]{\hat{R}_{#1}(p)}
\newcommand{\DDR}[2]{\hat{R}^{#1}_{#2}(p)}
\newcommand{\DDDR}[2]{\hat{R}^{#1}_{#2}(p')}

\def\A{{A}}

\def\eup{\varepsilon^{|1  \dots n{\cal i}}}
\def\edo{\varepsilon_{{\cal h}1  \dots n|}}
\def\eupp{\varepsilon^{|1  \dots n{\cal i}}(p)}
\def\edop{\varepsilon_{{\cal h}1  \dots n|}(p)}

\def\eu2{\varepsilon^{|2  \dots n{+}1{\cal i}}}
\def\ed2{\varepsilon_{{\cal h}2  \dots n{+}1|}}
\def\eu2p{\varepsilon^{|2  \dots n{+}1{\cal i}}(p)}
\def\ed2p{\varepsilon_{{\cal h}2  \dots n{+}1|}(p)}

\def\Eup{ {\cal E}^{|1  \dots n{\cal i} } }
\def\Edo{ {\cal E}_{{\cal h}1  \dots n| } }
\def\Eupp{{\cal E}^{|1  \dots n{\cal i}}(p)}
\def\Edop{{\cal E}_{{\cal h}1  \dots n|}(p)}

\def\U2{ {\cal E}^{|2  \dots n{+}1{\cal i} } }
\def\D2{ {\cal E}_{{\cal h}2  \dots n{+}1| } }
\def\Eu2p{{\cal E}^{|2  \dots n{+}1{\cal i}}(p)}
\def\Ed2p{{\cal E}_{{\cal h}2  \dots n{+}1|}(p)}

\newcommand{\eupi}[1]{\varepsilon^{|#1 \dots n+ #1 -1{\cal i}}}
\newcommand{\edoi}[1]{\varepsilon_{{\cal h}#1 \dots n+ #1 -1|}}
\newcommand{\Eupi}[1]{{\cal E}^{|#1 \dots n+ #1 -1{\cal i}}(p)}
\newcommand{\Edoi}[1]{{\cal E}_{{\cal h}#1 \dots n+ #1 -1|}(p)}
\newcommand{\N}[2]{N^{| #1 {\cal i}}_{\;\;\; {\cal h} #2 |}}
\newcommand{\K}[2]{K^{| #1 {\cal i}}_{\;\;\;\;\;\;\;\: {\cal h} #2 |}}
\newcommand{\iK}[2]{{K^{-1}}^{| #1 {\cal i}}_{\;\;\; {\cal h} #2 |}}

\def\bbr{{\rm I}\!{\rm R}}
\def\bbz{{\rm Z}\!\!\!{\rm Z}}
\def\subbc{{\rm C}\kern-3.3pt\hbox{\vrule height4.8pt width0.4pt}\,}
\def\BbbZ{Z\!\!\! Z}
\def\BbbN{{\rm I}\!{\rm N}}

\def\qd{\stackrel{.}{q}}
\def\pl{\partial}
\def\ig{\iota ({\rm tr} (gX\frac{\partial}{\partial g}) )}
\hyphenation{di-men-si-o-nal}
\begin{center}

%%%%%%%%%%%%%%%%%%%%%%%%  T I T L E  %%%%%%%%%%%%%%%%%%%%%%%

%% {\Huge\bf Chiral zero modes}\\[3 mm]
%% {\Huge\bf of the $SU(n)\,$ WZNW model}\\[18 mm]
{\Large\bf Chiral zero modes of the $SU(n)\,$}\\[3 mm]
{\Large\bf Wess-Zumino-Novikov-Witten model}\\[18 mm]

%%%%%%%%%%%%%%%%%%%%%%%%%% AUTHORS %%%%%%%%%%%%%%%%%%%%%%%%%%

{\large{\bf
P. Furlan$^{a,b}$ \footnote[1]{e-mail address: furlan@trieste.infn.it},
L.K. Hadjiivanov$^{c,a,b}$ \footnote[2]{Permanent address:
Theoretical Physics Division, Institute for Nuclear Research
and Nuclear Energy,
%% Tsarigradsko Chaussee 72,
BG-1784 Sofia, Bulgaria,
e-mail: lhadji@inrne.bas.bg},
I.T. Todorov$^{c,d}$ \footnote[3]{
Permanent address: Theoretical Physics Division, Institute for Nuclear
Research and Nuclear Energy,
%% Tsarigradsko Chaussee 72,
BG-1784 Sofia, Bulgaria, e-mail: todorov@inrne.bas.bg}\\}} \vskip 0.5 cm
$^a$Dipartimento di Fisica
Teorica dell' Universit\`a degli Studi\\ di
Trieste, Strada Costiera 11, I-34014 Trieste, Italy\\
$^b$Istituto Nazionale di Fisica Nucleare (INFN),
Sezione di Trieste,\\
Trieste, Italy\\
$^c$Theoretical Physics Division, Institute for Nuclear Research
and\\ Nuclear Energy, Tsarigradsko Chaussee
72, BG-1784 Sofia, Bulgaria\\
$^d$ Erwin Schr\"odinger International Institute for Mathematical Physics,
Boltzmanngasse 9, A-1090 Wien, Austria\\

\end{center}

\vspace{1cm}
{\bf MSC codes: }

\noindent
17B37 Quantum groups (quantum enveloping algebras) and related
deformations

\noindent
53D05 Symplectic manifolds

\noindent
81S10 Geometry and quantization, symplectic methods

\vspace{6mm}
{\bf Keywords:}

\noindent
WZ term, classical YBE, monodromy matrix, quantum exchange relations

%% {\bf To Paolo: HAPPY BIRTHDAY!!!}
%% \date{today}
\newpage
%%%%%%%%%%%%%%%%%%%%%% ABSTRACT %%%%%%%%%%%%%%%%%%%%%%%%%%%%

\begin{abstract}

%% {\normalsize
\noindent
We define the chiral zero modes' phase space of the
$G=SU(n)\,$ Wess-Zumino-Novikov-Witten (WZNW) model as an
$(n-1)(n+2)$-dimensional
%% remark: (n-1)(n+2) = n^2 +n -2 = (n^2 - 1) + (n-1)
%% (a [fix det], p [P=0])
manifold ${\cal M}_q\,$ equipped with a symplectic form $\O_q\,$ involving
a Wess-Zumino (WZ) term $\rho\,$ which depends on the monodromy $M\,$ and is
implicitly defined (on an open dense neighbourhood of the group unit) by
$$ d \r (M) = \frac{1}{3}\, {\rm {tr}}\, (M^{-1} d M)^3\,. \eqno{(*)}$$
This classical system exhibits a Poisson-Lie symmetry that evolves upon
quantization into an $U_q(s\ell_n )\,$ symmetry for $q\,$
a primitive even root
of $1\,.$ For each (non-degenerate, constant) solution of the
classical Yang-Baxter equation 
%% (CYBE) 
we write down explicitly a $\r (M)\,$ 
satisfying Eq.$(*)$ and invert the form $\O_q\,,$ thus
computing the Poisson bivector of the system. The resulting
Poisson brackets (PB) appear as the classical counterpart of the
exchange relations of the quantum matrix algebra studied
previously in \cite{FHIOPT}.
We argue that it is advantageous to equate the determinant $D\,$
of the zero modes' matrix $(a^j_\a )\,$ to a pseudoinvariant
under permutations $q$-polynomial in the $SU(n)\,$ weights, rather
than to adopt the familiar convention $D = 1\,.$
A finite dimensional "Fock space" operator realization of the factor
algebra ${\cal M}_q/{\cal I}_h\,,$ where ${\cal I}_h\,$ is an appropriate
ideal in ${\cal M}_q\,$ for $q^h = -1\,,$ is briefly discussed.

\end{abstract}

\newpage

%%%%%%%%%%%%%%% I N T R O D U C T I O N %%%%%%%%%%%%%%%%%

\textwidth = 16truecm \textheight = 22truecm \hoffset = -1truecm
\voffset = -2truecm

\section{Introduction}

\setcounter{equation}{0}
\renewcommand\theequation{\thesection.\arabic{equation}}

Two-dimensional conformal current algebra models are known to lead to an 
unconventional problem of classical symplectic dynamics defined in terms of a 
multivalued action \cite{WZ, N, W}
%% [47] (N.B.: since we are praised for providing good 
%% references shouldn't we also cite here Wess-Zumino and Nivikov?), 
or, equivalently, by a closed -- but not exact -- 3-form \cite{G},
%% [34], 
depending on a group valued field. It 
has been noted at an early stage of this development \cite{AF}
%% [1] 
that the most interesting new features of the theory already appear in a finite dimensional 
"toy model". The present paper is devoted to a study of a version of such a 
finite dimensional "chiral zero modes'" model. We display its precise relation 
to the (infinite dimensional) WZNW theory, reformulate it as a constrained 
dynamical system in the case when the underlying group is $SU(n)\,$, compute Poisson
brackets among the basic dynamical variables for a given non-degenerate solution 
of the classical (dynamical) Yang-Baxter equations and demonstrate that they 
appear as a (quasi)classical limit of quantum exchange relations considered 
earlier \cite{FHIOPT}.
%% [31].

\subsection{The zero modes' manifold}

\medskip

Let $G\,$ be a semisimple compact Lie group of $n\times n\,$
matrices with Lie algebra ${\cal G}\,.$ The zero modes' manifold
of a chiral WZNW model is not uniquely determined by the
corresponding $2$-dimensional ($2D\,$) conformal theory. It
depends on the splitting of the $G$-valued field $g(x^0,x^1)\,$
into chiral factors,
\be
g(x^0,x^1) = g_L (x^1+x^0) g_R^{-1} (x^1-x^0)\,,
\lb{gLR}
\ee
which obey a twisted periodicity condition (involving
monodromy degrees of freedom),
\be
g_C(x+2\pi ) = g_C(x) M\,,\quad C=L,R\,,\quad M\in G\,,
\lb{M}
\ee
implying that the $2D$ field is
periodic: $g(x^0,x^1+2\pi ) = g(x^0,x^1)\,.$ A further
arbitrariness is involved in the factorization of the chiral
fields $g_C (x)\,$ into (classical counterparts of) {\em chiral
vertex operators} $u(x)\,$ and {\em zero modes} $a\,;$
we shall write, in particular, the left movers' field in the form
\be
g_L (x)^A_\a = u (x)^A_j\, a^j_\a\qquad (A, j, \a =
1,\dots ,n )\,.
\lb{g=ua}
\ee
The chiral vertex operators have, by
definition, diagonal monodromies so that the ($x$-independent)
matrix $a = (a^j_\a)\,$ is chosen to diagonalize $M\,:$ \be a M =
M_p a\,,\quad M_p = \bq^{2{\hat p}}\,,\quad q =
e^{-i\frac{\pi}{k}}\,,\ \ \bq = e^{i\frac{\pi}{k}}\,. \lb{Mp} \ee
Here $k\,$ is the {\em Kac-Moody level} appearing as a coupling
constant in the WZNW model \cite{W} and ${\hat p}\,$ is a diagonal
matrix whose entries define a weight vector belonging to the Weyl
alcove ${\cal A}_n\,$ of the dual to the Cartan subalgebra of
${\cal G}\,.$ For ${\cal G} = su(n)\,$ \be \hat p =
\left(\matrix{p_1&0&\dots&0\cr 0&p_2&\dots&0\cr
\dots&\dots&\dots&\dots\cr 0&0&\dots &p_n} \right) \lb{p} \ee and
the Weyl alcove can be conveniently identified with \be {\cal A}_n
= \left\{ p = \{p_i\}_{i=1}^n\,,\quad p_{ij}:=p_i-p_j
> 0 \ {\rm for}\
i<j\,,\ \ P:=\frac{1}{n}\sum_{i=1}^n p_i = 0\,\right\}\,,
\lb{Weylalc} \ee $p_j\,$ playing thus the role of barycentric
coordinates.

While the weights $p_{\a_j}\,$ corresponding to the simple roots
$\a_j\,$ of ${\cal G}\,$ ($p_{\a_j} = p_{j\, j+1}\,$ for ${\cal G}
= su(n)\,$) provide an intrinsic characteristic of the state space
of (both the chiral and the $2D$) WZNW model, the zero mode matrix
$a^j_\a\,$ is gauge dependent. We shall use this freedom to work
in a "covariant but not unitary gauge" (discussed in
Section 3 below) and to equate, for $G=SU(n)\,,$ the determinant
$D\,$ of $(a^j_\a )\,$ to a pseudo-invariant under permutations of
$p_j\,$ function of $p\,$ (cf. \cite{FHIOPT}),
\ba
\lb{D}
&&D := \det (a^j_\a ) = {\cal D}_q (p) := \prod_{i<j} [p_{ij} ]\quad
{\rm for}\ \ G=SU(n)\,,\\
&&[p]:=\frac{q^p - \bq^p}{q-\bq}\quad (\, q \bq\, =\, 1\, )\,,\nonumber
\ea
rather than to $1\,$ as done in most related
studies \cite{AF, AT, BDF, BF, BFP1, BFP2, DR}.

\vspace{5mm}

\noindent {\bf Remark 1.1~} We use on purpose different notation
for the indices like $A, j, \a\,$ of $u\,$ and $a\,$ that run in
the same range (\ref{g=ua}) since they have rather different
nature. While the chiral model is invariant under left shifts of
$G\,$ (acting on $A\,$), it only admits a Poisson-Lie (or quantum
group) symmetry with respect to $\a\,,$ while $j\,$ labels the
diagonal elements of $M_p\,.$

%% \vspace{5mm}

\subsection{The case $n=2\,$ and its $k\to\infty\,$ limit.
The form $\O_q\,$ for $SU(2)$}

The advantage of the Ansatz (\ref{D}) (as compared to the
conventional $D=1\,$) is exhibited on the simple example of the
$SU(2)\,$ model space and its $q$-deformation which we proceed to
sketch. It can be also viewed as an introduction to the general case.

The realization of all irreducible representations (IR) of
$SU(2)\,$ with multiplicity $1\,$ in the Fock space of a pair of
creation and annihilation operators is half a century old (see
\cite{Sch} and \cite{Ba}). Its classical counterpart is the space
${\C}^2\,$ regarded as a K\"ahler manifold with a symplectic form
\be
\O_1 = \left(\, i dz_\a \wedge d {\bar z}^\a \equiv\,\right)\,
i d z_\a d{\bar z}^\a \equiv i (d z_1 d {\bar z}^1 + d z_2 d {\bar z}^2 )\,.
\lb{On2}
\ee
(We omit throughout this paper the wedge
sign $\wedge\,$ for exterior product of differentials but keep it
for the skew product of vector fields.) The corresponding Poisson
bivector,
\be
{\cal P}_1 = i\, \frac{\partial}{\partial z_\a}
\wedge \frac{\partial}{\partial{\bar z}^\a}\,,
\lb{Pbi2}
\ee
yields the PB counterpart of the canonical commutation relations
for (boso\-nic) creation and annihilation operators: \be \{ z_1 ,
z_2 \} = 0 = \{{\bar z}^1, {\bar z}^2 \}\,,\quad \{ z_\a , {\bar
z}^\b \} = i \d^\b_\a\,. \lb{classCCR} \ee In order to express
$\O_1\,$ (\ref{On2}) in terms of the above "group like" variable
$a = (a^j_\a )\,$ and "weight" $p\equiv p_{12}\,,$ we set
\ba
\lb{ap1}
a = \left(\matrix{z_1&z_2\cr -{\bar z}^2&{\bar
z}^1}\right)\,,\quad p:= \det a = z_1 {\bar z}^1 + z_2 {\bar z}^2\
( > 0\ \Leftrightarrow\  p\in {\cal A}_2 )\,,\\
{\hat p} = \frac{1}{2}\, \s_3 p\,.
\quad\qquad\qquad\qquad\qquad\qquad\qquad
\lb{ap2}
\ea
A simple
calculation allows then to rewrite $\O_1\,$ as an exact $2$-form:
\be \O_1 = - i\, d\, {\rm tr}\, \left( \hat p\, d a a^{-1}
\right)\,. \lb{On2-2} \ee The symplectic form $\O_q\,$ for the
$SU(2)_k\,$ WZNW zero modes (derived for the general $SU(n)_k\,$
case in Section 2 below) appears as a $1$-parameter deformation of
(\ref{On2-2}): \be \O_q (a, M_p) = \frac{k}{4\pi}\,\{{\rm tr}\,
(da a^{-1} ( 2 d M_p M_p^{-1} + M_p da a^{-1} M_p^{-1} ) ) - \r
(a^{-1} M_p a )\,\}\,. \lb{On2q} \ee Here $M_p\,$ is the diagonal
matrix defined in (\ref{Mp}), and $\r\,$ is
the WZ term:
\be
\frac{k}{2\pi} d M_p M_p^{-1} = i\, d \hat
p\,,\quad d \r (M) = \frac{1}{3}\, {\rm tr}\, (d M M^{-1} )^3\,.
\lb{rho-n2}
\ee
(The $3$-form in the right hand side is closed but not exact on
$G\,;$ the complex $2$-form $\r\,$ can only be defined on an open
dense neighbourhood $G_0\,$ of the identity of $G\,.$)

The phase space ${\cal M}_q\,$ is a $4$-dimensional surface in the
$5$-dimensional space of variables $a^j_\a\,$ and $p\,$, singled
out by the equation (\ref{D}): \be (\det a \equiv )\ D =
[p]\quad (\,\to\, p\ \ {\rm for}\ \ k\to\infty\,, \ {\rm resp.}\ q\to 1 )\,.
\lb{D2} \ee To summarize: for (undeformed) $SU(2)\,$ creation and
annihilation operators the determinant (\ref{ap1}) plays the role
of a number operator. More precisely, in the quantum theory
$p\in{\Bbb N}\,$ is the dimension of the IR of $SU(2)\,$ spanned
by all homogeneous polynomials of the creation operators
$a^1_\a\,$ of degree $p-1\,$ (acting on the Fock space vacuum).
For $p>0\,$ we can introduce new matrix variables with determinant
$1\,,$ \be g^j_\a := \frac{1}{\sqrt{p}} a^j_\a\,,\quad \det
(g^j_\a ) = 1 \lb{ga} \ee preserving the form of $\O_1\,$ ($ = - i
d {\rm tr}\, (\hat p d g g^{-1} )$). The new variables $( g^j_\a
)\,$ obeying (\ref{ga}), however, would not satisfy the canonical
PB relations for creation and annihilation operators. For $q\ne
1\,$ ($k\,$ finite) a change of variables $a^j_\a\ \to \ g^j_\a =
[p]^{-1/2} a^j_\a\,$ (that would again give $\det a = 1\,$) may
become singular, as $[k]=0\,$ for $q\,$ given by (\ref{Mp}). From
this point of view, the convention $\det a = 1\,$ is neither
convenient nor always possible.

\subsection{Outlook and references}

Although the WZNW model was introduced \cite{W} in terms of a
multivalued action, its solution was first given in the axiomatic
approach to conformal current algebra models \cite{KZ, T}. The
canonical (Lagrangean) approach had to wait the discovery of the
link between the quantum exchange relations and the Yang-Baxter
equation \cite{B}. It was initiated for the WZNW model in
\cite{BB} and was given a strong impetus by \cite{F1}. Among
early subsequent work (\cite{BDF, B2, G, ChuG1, ChuG2, FG, BBB,
FHT2, FHT3}) we would like to single out the development by
Gaw\c{e}dzki and coworkers \cite{G, FG, GTT} of a truly canonical
first order formalism adapted to the problem.
The present paper is devoted to a self-contained study of the finite
dimensional zero modes' problem (without recurrent appeal to its infinite
dimensional origin). This problem was first singled out in
\cite{AF} followed by \cite{AT, BF, FehG} --  among others.
It has an interest of its own, exhibiting in a nutshell a number of
properties that attract the attention of both physicists and
mathematicians: Poisson-Lie symmetry \cite{STS, AT, BFP1, BFP2, AKM,
BFP3},
$r$- ($R$-) matrices (classical and quantum) \cite{BD, STS, FRT},
dynamical $r$- ($R$-) matrices \cite{GN, Felder, EV, LX, L, ES, E, AM}.
The study of the $SU(2)\,$ case in \cite{FHT2} was extended to
$SU(n)\,$ in \cite{HIOPT} and \cite{FHIOPT}, $s\ell (n)\,$ being
singled out among other simple Lie algebras by the fact that the
corresponding quantum $R$-matrices satisfy quadratic (Hecke
algebra) relations. The gauge freedom in the very definition of
the zero mode phase space was discussed in \cite{FHT3} and its BRS
(co)homology was studied in \cite{DVT1, DVT2} (for a concise
review -- see \cite{goslar}). The presence of such a freedom
allows, in particular, to avoid the complications of the unitary gauge
advocated in \cite{BFP1, BFP2}.

\subsection{Outline of the paper}

After sketching (in Section 2.1) the derivation of the expression
(\ref{Onnq}) for $\O_q\,$ that generalizes (\ref{On2q}) to any
compact semisimple Lie group $G\,,$ we study in Section 2.2 an
extension ${\cal M}_q^{ex}\,$ of the phase space ${\cal M}_q\,$
for $G=SU(n)\,$ for which one derives a more manageable
symplectic form $\O_q^{ex}\,.$ In Section 2.3 we display the
undeformed limit $k\to\infty\ (q\to 1 )\,$ in which the WZ term
disappears. The resulting form $\O_1\,$
can be easily inverted. We also display the
%% momentum maps
Hamiltonian vector fields corresponding to the constraints
$\chi := {\rm log} \frac{D}{{\cal D}_q (p)}\,$
and $P := \frac{1}{n} \sum_{s=1}^n p_s\,.$
In particular,
\be
i\frac{\hat{\p}}{\p P} \O_q^{ex} = i \sum_{s=1}^n
\frac{\hat{\p}}{\p p_s} \O_q^{ex} = d\chi = \frac{d D}{D} -
\frac{d {\cal D}_q (p)}{{\cal D}_q (p)}\,.
\lb{chi}
\ee
Here ${\hat X} \O\,$ means the contraction of the
vector field $X\,$with the form $\O\,;$ we have, e.g.,
\be
\frac{\hat{\p}}{\p p_s} d p_j = \d^s_j - d p_j \frac{\hat{\p}}{\p p_s}\,.
\lb{contrps}
\ee
It is important that these "momentum maps"
remain valid after $q$-de\-for\-ma\-tion (i.e., for finite $k\,$).
Section 3 is devoted to inverting the form $\O_q^{ex}\,$ (and $\O_q\,$),
thus computing PB among zero modes.
In Section 4.1 we demonstrate that the quasiclassical limit ($k \gg n\,,\
p_{j\ell} \gg 1\,,\ \frac{p_{j\ell}}{k}\,$ finite) of the quantum exchange
relations of
\cite{HIOPT, FHIOPT, HST} reproduces the PB relations of Section 3.
In the rest of Section 4 we review the $U_q(s\ell_n )\,$ symmetry of the
quantum matrix algebra and its operator realization.

\section{Zero modes' phase space from chiral\\ WZNW $2$-form}

\setcounter{equation}{0}
\renewcommand\theequation{\thesection.\arabic{equation}}

\subsection{From $2D\,$ canonical $3$-form to zero modes'
symplectic form}

The canonical approach to a field theory in $D$-dimensional
space-time formulated in \cite{G} (where its sources are cited and
reviewed) starts with a closed $(D+1)$-form $\o\ ( \, = d {\bf L}
(x)\,$ if a Lagrangean $D$-form ${\bf L} (x)\,$ exists). It allows
to read off the equations of motion while the integral over a
$(D-1)$-dimensional space-like surface provides the symplectic
form of the theory. A form of this type, called symplectic
density, was recently (partly rediscovered and) applied to
Yang-Mills, general relativity, Chern-Simons and supergravity
theories \cite{JS}. In the case of the WZNW model the $3$-form
$\o\,$ can be written as the sum of an exact form and
%% a WZ term
%% the normalized canonical bi-invariant closed $3$-form on the group
%% $G\,$ (multiplied by a positive integer $k\,$),
the canonical invariant closed $3$-form on the group $G\,,$
\be
\o = d\,\left\{ \frac{1}{2}\,{\rm tr}\,\left( i g^{-1} d g +
\frac{\pi}{k} {\bf J}\,\right)\, \sbJ \, \right\} +
\frac{k}{12\pi}\,{\rm tr}\, (g^{-1} d g )^3\,,
\lb{omega}
\ee
where $\bJ\,$ is the current $1$-form and $\sbJ\,$ is its Hodge
dual: \be {\bJ} (x) = j_\mu (x) d x^\mu\,,\ \ \sbJ (x) =
\e_{\mu\nu} j^\mu (x) d x^\nu\quad (\,\e_{\mu\nu} = -
\e_{\nu\mu}\,,\ \e_{01} = 1 = \e^{10}\, )\,. \lb{J*J} \ee
We shall
sum up without derivation the implications of Eq.(\ref{omega}).

The equations of motion, obtained as the pull-back of the
contractions of $\o\,$ with the vertical vector fields
$\frac{\d}{\d j_\mu (x)}\,$ and $g(x) X \frac{\d}{\d g (x)}\,,$ read
\be \bJ = \frac{k}{2\pi i}\, g^{-1} d g\,,\ \ d \bJ + \frac{2\pi
i}{k} \bJ^2 = 0\ \ \Rightarrow\ \ d (\bJ + {^*\!{{\bf J}}} ) = 0\,.
\lb{eqmot} \ee They imply the existence of left and right
(N\"other) currents depending on a single light cone variable, \be
j_R = \frac{1}{2} (j^0+j^1) ,\  j_L=\frac{1}{2} g (j^1 - j^0 )
g^{-1}\,,\ \ \p_+ j_R = 0 = \p_- j_L\ \ {\rm for}\ \ \p_\pm =
\frac{1}{2} (\p_1\pm\p_0 )\,, \lb{jLR} \ee and the factorization
(\ref{gLR}) of $g(x^0,x^1)\,.$

The symplectic form $\O^{(2)}\,$ can be expressed in terms of
either of the two chiral currents: \ba \lb{O2}
&&\O^{(2)} = \int^{\pi}_{-\pi}\,\o\, dx^1 =\\
&&= - \int_{-\pi}^{\pi}\, dx\,{\rm tr}\, \left( i d (j_L dg
g^{-1}) + \frac{k}{4\pi} dg g^{-1} (dg g^{-1} )' \right) = \nonumber\\
&&= \int_{-\pi}^{\pi}\, dx\,{\rm tr}\, \left( i d (j_R g^{-1} dg )
+ \frac{k}{4\pi} g^{-1} dg (g^{-1} dg)' \right)\,. \nonumber \ea
Inserting the factorized expression (\ref{gLR}) for $g\,$ in
(\ref{O2}), one can split $\O^{(2)}\,$ into chiral symplectic
forms
\be
\O^{(2)} = \O
(g_L , M) - \O (g_R , M)\,, \lb{OLR} \ee where \be \O (g_C ,M ) =
\frac{k}{4\pi}\left\{ {\rm tr}\left( \int_{-\pi}^\pi dx (g_C^{-1}
d g_C (g_C^{-1} d g_C )' ) + b_C^{-1} d b_C d M M^{-1} \right) -
\r (M) \right\}
\lb{OC}
\ee
with
\be
b_C := g_C (-\pi )\,,\quad M = b_C^{-1} g_C (\pi )\ \,
\left(\, = g_C^{-1} (x) g_C (x+2\pi )\,\right)\,,\ \ C=L,R\,.
\lb{bM}
\ee
The cumbersome (ill defined) WZ term $\r (M)\,$ (satisfying
(\ref{rho-n2})) has been added and subtracted
from the two chiral terms to ensure $d \O\, =\, 0\,.$ An
alternative approach, introducing quasi-Poisson manifolds
\cite{AKM} (for which the Jacobi identity satisfied by proper PB is
replaced by a weaker condition) is developed in \cite{BFP3}.

Finally, substituting $g_L (x)\,$ by its expression (\ref{g=ua}),
we find
\be
\O (g_L , M) = \O (u , M_p) + \o_q (M_p) + \O_q (a ,
M_p )
\lb{Oua}
\ee where
\ba
\lb{Onnq}
\O_q (a, M_p)&=& \\
= \frac{k}{4\pi}\,\{{\rm tr}\, (da a^{-1} ( 2 d M_p M_p^{-1} &+&
M_p da a^{-1} M_p^{-1} ) ) - \r (a^{-1} M_p a )\,\} - \o_q(M_p)
\nonumber
\ea and $\o_q\,$ is an arbitrary closed $2$-form (which
will be restricted further by some symmetry conditions). For
$G=SU(2)\,$ there is a single variable $p\,,$ hence $\o_q
(M_p)\equiv 0\,$ and (\ref{Onnq}) coincides with (\ref{On2q}).

A detailed derivation of the results formulated in this subsection
will be presented elsewhere.

\subsection{Basis of right invariant $1$-forms. An extended pha\-se
space and a privileged choice of $\o_q\,$ for $G=SU(n)$}

We shall now write down the first two terms in the expression
(\ref{Onnq}) as sums of products of right invariant forms. To
this end we shall use the Cartan-Weyl basis $\{ h_i , e_\a \}\,,\
\a\,$ running through the positive roots of ${\cal G}_{\subbc}\,$
(in its
$n$-dimensional fundamental representation) satisfying \be [ h_i ,
h_j ] = 0\,,\ \  [h_i , e_{\pm\a} ] = \pm 2 \frac{(\a | \a_j
)}{|\a_j |^2} e_{\pm\a}\,,\ \ [e_{\a_i}, e_{-\a_j} ] = \d_{ij} h_j
\lb{Cartan-Weyl} \ee ($i,j = 1,\dots ,r:=\, {\rm rank}\,{\cal
G}\,$), and shall write
\be
\hat p = \sum_{j=1}^r p_{\a_j}
h^j\quad{\rm with}\ \ {\rm tr}\, (h^i h_j ) = \d^i_j\quad
(\,{\rm and}\ \,{\rm tr}\,(e_\a e_{-\b}) = \d_{\a\b}\, )
\lb{tr}
\ee
(thus $\{ h_i\}\,$ and $\{ h^j\}\,$
define dual bases of diagonal matrices).
Let further $\Theta^j\,,\ \Theta^{\pm\a}\,$ and $\frac{d D}{D}\,$
be the corresponding right invariant $1$-forms in $T^*
G_{\subbc}^{ex}\,,$
\be
G_{\subbc}^{ex} := (G \times {\Bbb R}_+ )_{\subbc} \,,
\lb{Gex}
\ee
defined by
\be
\Theta^j = -i\,{\rm tr}\,
(a^{-1} h^j da )\,,\quad \Theta^{\pm\a} = -i\,{\rm tr}\, (a^{-1}
e_{\mp\a} da )\,,\quad \frac{dD}{D} = {\rm tr}\, (da a^{-1})\,.
\lb{TD}
\ee
It then follows that
\be
- i d a a^{-1} = \sum_{j=1}^r
\Theta^j h_j + \sum_{\a >0} (\Theta^\a e_\a + \Theta^{-\a} e_{-a}
) - \frac{i}{n} \frac{dD}{D} \id
\lb{daa-1}
\ee
where $D = \det a > 0\,$ and $\id\,$ is the $n\times n\,$ unit matrix.

If $G\,$ is compact, then the forms
$\Theta^j\,$ are real while
$\Theta^{-\a}\,$ are complex conjugate to $\Theta^\a\,.$ We also
note that the (Lie algebra valued) $1$-form (\ref{daa-1})
is not closed but
defines a flat connection, the $\Theta$'s satisfying the
Cartan-Maurer relations. We shall use, in particular, \be d
\Theta^j = i \sum_{\a >0} (\L^j | \a ) \Theta^\a
\Theta^{-\a}\quad\quad\ (\ ( \L^j | \a_\ell ) = \d^j_\ell \ )\,,
\lb{CM} \ee $\L^j\,$ being the {\em fundamental weights} of ${\cal
G}\,.$

Inserting (\ref{tr}) into the first term in the right hand side of
(\ref{Onnq}) and using (\ref{rho-n2}) and (\ref{TD}), we deduce
\be \frac{k}{2\pi}\,{\rm tr}\, (d a a^{-1} dM_p M_p^{-1} ) =
i\,{\rm tr}\, (d a a^{-1} \hat p ) = \sum_{j=1}^r d p_{\a_j}
\Theta^j\,.
\lb{1term}
\ee
The second term is expressed as a sum
of products of conjugate off-diagonal forms:
\be
\frac{k}{4\pi}\,{\rm tr} ( d a a^{-1} M_p da a^{-1} M_p^{-1} ) =
\frac{k}{4\pi}\,(\bq -q) \sum_{\a >0} [2 p_\a ] \Theta^\a
\Theta^{-\a}\,,
\lb{2term}
\ee
where $p_\a\,$ is a linear
functional on the roots: \be p_\a = \sum_{j=1}^r (\L^j | \a )
p_{\a_j}\quad{\rm for}\quad \a = \sum_{j=1}^r (\L^j | \a ) \a_j\,;
\lb{pa} \ee here $(\L^j | \a )\in \Z_+\,$ and we have the relation
\be Ad_{M_p} e_\a := M_p e_\a M_p^{-1} = \bq^{2p_\a} e_\a\,.
\lb{AdMp} \ee

At this point we shall specialize to the case $G=SU(n)\,$ and
will view the $(n-1)(n+2)$-dimensional symplectic manifold
${\cal M}_q = {\cal M}_q (n)\,$ as a submanifold
of codimension $2\,$ in the {\em
extended} ($n(n+1)$-dimensional) {\em phase space} ${\cal
M}_q^{ex}\,$ spanned by $p_i\,$ and $a^j_\a\ (i,j,\a = 1,\dots ,n
)\,$ regarded as independent variables:
\be
{\cal M}_q = \left\{ (p_i , a^j_\a ) \in {\cal M}_q^{ex}\,;
\ P:=\frac{1}{n}\sum_{s=1}^n p_s = 0\,, \
\chi := {\rm log} \frac{D}{{\cal D}_q (p)} = 0\,\right\}\,.
\lb{Mq}
\ee
We introduce the Weyl basis $\{ e^j_i \}\,$ of $n\times n\,$
matrices satisfying
\be
e^j_i\, e^\ell_k \, = \, \d^j_k \, e^\ell_i\,,\quad
(e^j_i )^\ell_{~k} = \d^\ell_i \d^j_k\,,\ \
i,j,k,\ell = 1,\dots ,n\,.
\lb{eij}
\ee
The positive roots
$\a_{ij}\ (i<j)\,$ of $su(n)\,$ correspond to raising operators,
$e^j_i\,,$ while $- \a_{ij}\,$ are associated with lowering ones,
$e_j^i\,.$ Eq.(\ref{daa-1}) now assumes a simple explicit form:
\be
- i da a^{-1} = \Theta^j_k e^k_j\ \left(\, \equiv\,
\sum_{j,k=1}^n  \Theta^j_k e^k_j\, \right)\,,\quad \Theta^j_k = -i
{\rm tr}\, \left( e^j_k d a a^{-1} \right) = - i d a^j_\s
(a^{-1})^\s_k\,.
\lb{daa-1*}
\ee
The general Cartan-Maurer
relations (which incorporate (\ref{CM})) are written simply as \be
d \Theta^j_k = i \,\Theta^j_s \Theta^s_k\,. \lb{genCM} \ee

Recalling that the relation (\ref{D}) is invariant under
simultaneous permutation of the rows of the matrix $(a^j_\a )\,$
and of $p_j\,$ (i.e., under the action on both sides of the
$su(n)\,$ Weyl group), we shall also require permutation
invariance of the extended form $\o^{ex}_q (p)\,.$ We shall
determine $\o_q (M_p) =\o^{ex}_q (p) \mid_{P=0}\,$ by
further demanding that
the symplectic form $\O^{ex}_q\,$ on ${\cal M}^{ex}_q\,,$
\be
\O^{ex}_q = \sum_{s=1}^n d p_s \Theta^s_s - \frac{k}{4\pi}\left\{
(q-\bq ) \sum_{j<\ell} [2 p_{j\ell} ] \Theta^j_\ell \Theta^\ell_j
+ \rho (a^{-1} M_p a )\,\right\} - \o^{ex}_q (p)
\lb{Oqex}
\ee
will reduce to $\O_q\,$ on the surface ${\cal M}_q \subset {\cal
M}_q^{ex}\,.$ In order to implement this last condition, we shall
require that the terms involving $dP\,$ cancel in the difference
\be
- \o_q (M_p ) = \left( - i d P \frac{d{\cal D}_q (p) }{{\cal
D}_q (p)} \right) - \o_q^{ex} (p)\,.
\lb{oq}
\ee
Inserting the expression (cf. (\ref{D}))
for ${\cal D}_q (p)\,$ which implies
\be
\frac{d{\cal D}_q (p) }{{\cal D}_q
(p)} = \frac{\pi}{k} \sum_{j<\ell} {\rm cotg} ( \frac{\pi}{k}
p_{j\ell}) d p_{j\ell}\,,
\lb{dlogDp}
\ee
we find a form $\o_q^{ex} (p)\,$ satisfying all above conditions:
\be
\o_q^{ex} (p)\, = i \frac{\pi}{k} \sum_{j<\ell} {\rm cotg} (
\frac{\pi}{k} p_{j\ell}) d p_j d p_\ell\,.
\lb{oqex}
\ee
Indeed,
using the relation \be p_j = P + \frac{1}{n} \sum_{s=1}^n
p_{js}\,, \lb{pP} \ee we deduce \be \o_q (M_p ) = \frac{i\pi}{nk}
\sum_{1\le j<\ell <m\le n} \left( {\rm cotg} ( \frac{\pi}{k}
p_{j\ell}) + {\rm cotg} ( \frac{\pi}{k} p_{\ell m}) - {\rm cotg} (
\frac{\pi}{k} p_{jm}) \right) d p_{j\ell}\, d p_{\ell m} \lb{oq*}
\ee (note that for $n=2\,$ there is no triple $j,\ell, m\,$
satisfying the above inequalities so that the form $\o_q (M_p )\,$
vanishes, as it should, while $\o_q^{ex} (p)\,$ (\ref{oqex})
reduces to a single term: $\o_q^{ex} (p)\, =\, i \,
\frac{\pi}{k}\,  {\rm cotg}\, ( \frac{\pi}{k} p_{12})\, d p_1\, d
p_2\,$).

We observe the relative simplicity of the extended symplectic form
(\ref{Oqex}), (\ref{oqex}) as compared with $\O_q\,$ (obtained
from (\ref{Onnq}) by inserting (\ref{1term}) with \be p_{\a_j} =
p_{j\, j+1}\,,\quad \Theta^j = \frac{n-j}{n} \sum_{s=1}^j
\Theta^s_s - \frac{j}{n} \sum_{s=j+1}^n \Theta^s_s\,, \lb{pT} \ee
(\ref{2term}) and (\ref{oq*})). It is, therefore, rewarding to
know that the PB we are interested in can be computed using the
simpler expression $\O_q^{ex}\,,$ as we shall see in Section 3. In
the next subsection we shall display this property for the $k\to
\infty\,$ limit theory.

\subsection{Right invariant vector fields. The limit $k\to \infty\,$.
Dirac brackets}

It is easy to display the basis of right invariant vector fields
$\left\{ \frac{\p}{\p p_\ell} , V^k_j \right\}\,$ dual to the
basis $\left\{ dp_\ell , \Theta^j_k \right\}\,$ of $1$-forms:
\be
V^k_j = i\,{\rm tr}\,\left( e^k_j a \frac{\p}{\p a} \right) = i
a^k_\s \frac{\p}{\p a^j_\s }\,.
\lb{V}
\ee
Indeed, contracting the
form $\Theta^\ell_m\,$ (\ref{daa-1*}) with $V^k_j\,,$ we find
\be
{\hat V}^k_j \Theta^\ell_m = {\rm tr}\, (e^k_j a a^{-1} e^\ell_m ) =
\d^\ell_j \d^k_m\,,\qquad {\hat V}^k_j d p_\ell = 0\,;
\lb{VT}
\ee
obviously,
$$\frac{{\hat \p}}{\p p_j} \Theta^\ell_m = 0\,,
\qquad \frac{{\hat \p}}{\p p_j} d p_\ell =
\d^j_\ell\,.$$ This would allow to invert the form $\O_q^{ex}\,$
but for the WZ term.

We shall profit from the above remark taking the limit
$k\to\infty\,$ in which the WZ term disappears. Indeed, using the
expression for $q\,$ in (\ref{Mp}), we find
\be
\lim_{k\to\infty} \frac{k}{2\pi} (\bq -q) = i\,,\qquad
\frac{1}{2} \lim_{k\to\infty} [2p] = p
\lb{lim}
\ee
and hence,
\ba
&&\O_1^{ex} (a,p) =
\sum_{s=1}^n d p_s \Theta^s_s + i \sum_{1\le j<\ell\le n}
p_{j\ell} \Theta^j_\ell \Theta^\ell_j - \o_1 (p) =\nonumber\\
&&= d \sum_{s=1}^n p_s \Theta^s_s - i \sum_{1\le j<\ell\le n}
\frac{d p_j d p_\ell}{p_{j\ell}}\,.
\lb{O1ex}
\ea
Here we have set
\be
\lim_{k\to\infty} \frac{k}{4\pi} \rho (a^{-1} M_p a ) = 0\,.
\lb{rholim}
\ee
In fact, since the right hand side of (\ref{O1ex})
is a closed $2$-form, it follows that
\be
\lim_{k\to\infty}
\frac{k}{4\pi} {\rm tr}\, (d M M^{-1} )^3 = 0\quad {\rm for}\quad
M = a^{-1} M_p a\,;
\lb{limWZ}
\ee
we conclude that
$\frac{k}{4\pi} \rho\,$ can be also chosen to vanish in this limit
-- a property that can be derived from the expression for $\rho
(a^{-1} M_p a )\,$ given in Section 3.

As anticipated, it is straightforward to invert the $2$-form
(\ref{O1ex}). The result can be encoded in the Poisson bivector
\be
{\cal P} = \sum_{s=1}^n V^s_s \wedge \frac{\p}{\p p_s} + i
\sum_{1\le j<\ell\le n} \frac{1}{p_{j\ell}} ( V^\ell_j \wedge
V^j_\ell - V^j_j \wedge V^\ell_\ell\, )\,,
\lb{Pbiv}
\ee
which gives
rise to the following PB:
\be
\{ p_j , p_\ell \} = 0\,,\qquad \{ a^j_\a , p_\ell \} =
i\, \d^j_\ell\, a^j_\a
\lb{PB1-1}
\ee
and
\be
\{ a^j_\a , a^\ell_\b \} = r^{(1)} (p)^{j\ell}_{j'{\ell}'}\,
a^{j'}_\a \, a^{{\ell}'}_\b \qquad(\,{\rm i.e.,}\ \{ a_1, a_2 \} =
r^{(1)}_{12} (p) a_1 a_2 \ )
\lb{PB1-2}
\ee
where the undeformed
classical dynamical $r$-matrix is given by
\be
\lb{r1p}
r^{(1)}
(p)^{j\ell}_{j'{\ell}'}= \left\{
\begin{array}{ll}
\, \frac{i}{p_{j\ell}}\, (\d^j_{j'} \d^\ell_{{\ell}'} -
\d^j_{{\ell}'} \d^\ell_{j'} )\ & {\rm for}\ j\ne\ell\\
\, 0& {\rm for}\ j=\ell
\end{array}
\right. \,.
\ee

For a general Poisson manifold ${\cal M}\,$ with a pair of second
class constraints $P\,$ and $\chi\,$ the Dirac brackets $\{ f , g
\}_D\,$ \cite{D} of two arbitrary functions on ${\cal M}\,$ are expressed
in terms of their PB as
\be
\{ f , g \}_D = \{ f , g \} + \frac{1}{\{ P , \chi \}}
\left( \{ f , P \} \{ \chi, g \} - \{ f , \chi \} \{ P , g \} \right)\,.
\lb{Dbracket}
\ee
We shall verify that in the case at hand
\be
\{ p_{j\ell} , P \} = 0 = \{ p_{j\ell} , \chi \} \,,\quad \{
a^j_\a , \chi \} = 0\,.
\lb{PBmain}
\ee
The first pair of equations implies that $p_{j\ell}\,$ are
"observables" on ${\cal M}_q \subset {\cal M}_q^{ex}\,,$ so that
$\{ p_{j\ell} , f \}_D = \{ p_{j\ell} , f \}\,$
for any function $f\,$ on ${\cal M}_q^{ex}\,;$ in particular,
\be
\{ p_{j\ell} , a^m_\a \} = i (\d^m_\ell - \d^m_j ) a^m_\a =
\{ p_{j\ell} , a^m_\a \}_D\,.
\lb{pjl-a}
\ee
The last equation (\ref{PBmain}) is sufficient to assert that the
PB (\ref{PB1-2}) coincide with the corresponding Dirac brackets.

Although it is easy to verify (\ref{PBmain}) directly, using
(\ref{Pbiv})-(\ref{PB1-2}), we shall give a more general
derivation that will apply to the case of finite $k\ (q\ne 1)\,$
as well. To this end we shall use the momentum maps
\be
i \frac{{\hat \p}}{\p P} \O_1 = i \sum_{s=1}^n
\frac{{\hat \p}}{\p p_s} \O_1 = i \sum_{s=1}^n \Theta^s_s -
\sum_{1\le j<\ell\le n} \frac{d p_{j\ell}}{p_{j\ell}} = d \chi\,;
\ \ - \frac{1}{n} \sum_{s=1}^n {\hat V}^s_s \,\O_1 = d P\,.
\lb{momentummaps}
\ee
Displaying the Hamiltonian vector fields corresponding to
$\chi\,$ and $P\,,$ Eq. (\ref{momentummaps}) allows to compute
any PB of the constraints; in particular,
\ba
&&\{ \chi , a^j_\a \} = i \frac{{\hat \p}}{\p P}\, d a^j_\a = 0\,,\ \
\{ \chi , p_{j\ell} \} = i \frac{{\hat \p}}{\p P}\, d p_{j\ell}
= 0\,,\nonumber\\
&& \{ p_{j\ell} , P \} = \frac{1}{n} \sum_{s=1}^n
{\hat V}^s_s\, d p_{j\ell}  = 0\,,\qquad \{ P , \chi \} = -i\,.
\lb{constr-ap}
\ea
We find, on the other hand,
\be
\{ a^j_\a , p_\ell \}_D = \{ a^j_\a , p_\ell \} +
i \{ a^j_\a , P \} \{ \chi , p_\ell \}
= i a^j_\a (\d^j_\ell - \frac{1}{n} )\,.
\lb{apD}
\ee

\section{Inverting $\O_q^{ex}\,.$ PB in ${\cal M}_q (n)$}

\setcounter{equation}{0}
\renewcommand\theequation{\thesection.\arabic{equation}}

\subsection{The WZ form}

It was Gaw\c{e}dzki \cite{G} (see also \cite{FG}) who introduced in the
early 1990'ies the WZ $2$-form $\rho (M)\,$ and described its relation to
the non-degenerate (constant) solutions of the 
classical Yang-Baxter equation (CYBE). Gradually, a more
general and complete understanding of such a relation has been worked out
\cite{BFP2, FehG}. We shall only deal here with a special case of the
outcome of \cite{FehG} corresponding essentially to the early discussion
in \cite{FG}.

We shall again start with an arbitrary  semisimple matrix Lie group
$G\,$ with Lie algebra ${\cal G}\,.$ For an arbitrary pair $\{ t^a \}\,,\
\{ T_b \}\,$ of dual bases in ${\cal G}\,,$ we can write the Killing
metric tensor $\eta_{ab}\,$ and its inverse, $\eta^{ab}\,,$ as
\be
\eta_{ab} = {\rm tr}\, (T_a T_b )\,,\quad \eta^{ab} = {\rm tr}\, (t^a t^b )\quad{\rm for}\quad
{\rm tr}\, (t^a T_b ) = \d^a_b\,.
\lb{Killeta}
\ee
In the Cartan-Weyl basis $\{ T_a \} = \{ h_i , e_{\pm\a} \}\,$
we have $\{ t^a = h^i , e_{\mp\a} \}\,$
and the nonzero elements of $\eta\,$ are
\be
\eta_{ij} = {\rm tr}\, h_i h_j = (\a_i | \a_j )\,,
\quad \eta_{\a\b} = {\rm tr}\, e_\a e_\b = \d_{\a , -\b}
\lb{etanonzero}
\ee
(where the norm square of the highest root is fixed to $2\,$).

The polarized Casimir invariant $C_{12} \in {\rm Sym}\, ({\cal G}\otimes
{\cal G} )\,,$ given (in Faddeev's notation \cite{FRT}) by
\be
C_{12} = \eta_{ab}\, t_1^a\, t_2^b \ (\, =
T_a\otimes t^a\equiv t^a\otimes T_a\, )
= h_{i1}\, h^i_2 + \sum_\a e_{\a\, 1}\, e_{-\a\, 2}
\lb{Cas-Fadd}
\ee
where the sum is taken over all,
positive and negative, roots $\a\,,$ plays the role of the unit operator
on ${\cal G}\,:$
\be
C X := {\rm tr}_2\, ( C_{12} X_2 ) = X\, (\equiv X_1 )\quad{\rm for}\quad X\in {\cal G}\,.
\lb{Cas-unit}
\ee

Let $r_{12} = - r_{21}\, (\, \in {\cal G}\wedge {\cal G}\, )\,$ be a solution of the {\em modified} CYBE
\be
[r_{12} , r_{13} + r_{23} ] + [r_{13} , r_{23} ] =[ C_{12} , C_{23} ] \ \ ( = - f_{abc} t^a t^b t^c )\,,
\lb{MCYBE}
\ee
and let $r\,$ be the corresponding operator $( r:\, {\cal G} \to {\cal G}\,)\,$ defined by taking the trace in the second
argument as in (\ref{Cas-unit}):
\be
r X := {\rm tr}_2\, ( r_{12} X_2 ) \quad{\rm for}\ \ X\in {\cal G}\ \
\Rightarrow\ \ r_{12} = r\, C_{12}\,.
\lb{r-oper}
\ee

\noindent
{\bf Proposition 3.1~}
{\em Let the  $2$-form $\rho (M)\,$ be written in terms of a skew-symmetric kernel
$K(M)_{12} \in {\cal G}\wedge{\cal G}\,$ for $M\in G_0\,$ where
$G_0\,$ is an open dense neighbourhood of the group unit in which the operator
$(1- Ad_M ) r+ 1 + Ad_M\,,\ \ Ad_M X := M X M^{-1}\,,$ is invertible,
and let $K(M)\,$ be the corresponding operator
$K(M) : {\cal G}\to {\cal G}\,$ defined as in (\ref{r-oper}),
\be
\rho (M) = \frac{1}{2}\,{\rm tr}\, (dMM^{-1} K(M) dMM^{-1})\,, \quad \
( \, {\rm tr}\, (X K(M) X) = 0 \ \ \forall X\in {\cal G}\,).
\lb{defrhoK}
\ee
Assume further that
\be
K(M) = ((1+Ad_M)\, r + 1-Ad_M)((1-Ad_M)\, r + 1+Ad_M )^{-1}
\lb{KM}
\ee
so that $K(1) = r\,.$ Then $\rho (M)\,$ satisfies
%% (\ref{rho-n2}),
\be
d \r (M) = \frac{1}{3}\, {\rm tr}\, (d M M^{-1} )^3
\lb{rho-nN}
\ee
iff $r_{12}\,$ (related to $r\,$ by (\ref{r-oper}))
satisfies the modified CYBE (\ref{MCYBE}).}

\vspace{5mm}

The statement is a corollary of Propositions 1 and 2 of \cite{FehG}; see
also the earlier discussion in \cite{FG}.

\vspace{5mm}

\noindent
{\bf Remark 3.1~} The modified CYBE (\ref{MCYBE}) for
$r_{12}\,$ is equivalent to the standard CYBE
\be
[r^\pm_{12} , r^\pm_{13} + r^\pm_{23} ] + [r^\pm_{13} , r^\pm_{23} ] =
0\qquad {\rm for}\ \ r^\pm_{12} = r_{12} \pm C_{12}\,.
\lb{sCYBE}
\ee
In fact, Ref. \cite{FG} deals with Eq. (\ref{sCYBE}).

\vspace{5mm}

\noindent
{\bf Remark 3.2~}
Using (\ref{KM}), it is easy to check that the
skewsymmetry of $K (M)\,$ is equivalent to that of $r\,,\ \
{^t}K(M) = - K(M)\ \ \Leftrightarrow \ \ {^t}r = - r\,$ where
the transposition ${^t}\,$ is w.r. to the invariant bilinear form
tr.

\vspace{5mm}

\noindent
{\bf Remark 3.3~}
One can consider a more general Ansatz of type (\ref{KM}) allowing the
operator $r\,$ to depend on $M\,.$ Then one has to deal with a
"dynamical" version of the (modified) CYBE including
differentiation in the group parameters -- see Eqs. (1.3) and (3.8) of
\cite{FehG}. One argues in \cite{BFP2} that a constant classical
$r$-matrix cannot correspond to a compact group $G\,.$
It is well known, indeed, that the modified CYBE (\ref{MCYBE}) has
no real solution in ${\cal G}\wedge{\cal G}\,$ for ${\cal G}\,$ compact.
We shall however stick to the above simple choice which
uses a complex $2$-form $\rho\,.$
As noted in the Introduction, the use of the simple constant
$r$-matrix for the PB of the zero modes $a^j_\a\,$ is perfectly admissible
because the freedom in their choice does not affect the properties of
$u^A_j (x)\,$ which always transform covariantly under left shifts of the
compact group $G\,.$

\vspace{5mm}

Using (\ref{defrhoK}) and (\ref{KM}), we can present the WZ $2$-form
in (\ref{Onnq}) as
\be
\r (a^{-1} M_p a) = \frac{1}{2} {\rm tr}\, \{
( d M_p M_p^{-1} - A_- (da a^{-1})) K^a
( d M_p M_p^{-1} - A_- (da a^{-1})) \}\,.
\lb{rhoinform}
\ee
Here and below we are using the operators
\be
A_\pm := 1\pm Ad_{M_p}\,,\qquad
A_- dM_p M_p^{-1} = 0 = (A_+ - 2) dM_p M_p^{-1}\,,
\lb{Apm}
\ee
while $K^a\,$ is given by
\ba
&&K^a := Ad_a\, K(a^{-1} M_p a)\, Ad_a^{-1} =
(A_+ r^a + A_- ) (A_- r^a + A_+ )^{-1}\,,\nonumber\\
&&r^a := Ad_a \, r\, Ad_a^{-1}
\lb{Ka}
\ea
($K^a\,$ and $r^a\,$ are skewsymmetric together with $K(M)\,$ and $r\,$).

We note that $\r (a^{-1} M_p a)\,$ coincides with its extension on ${\cal M}_q^{ex}(n)\,.$ This
is obviously true for the $3$-form $d\r (M)\,$ (\ref{rho-nN}) for
$M=a^{-1} M_p a\,.$ Indeed,
the contribution of the term proportional to $dP\,$
is given by
\be
d \r^{ex} (a^{-1} M_p a) - d\r (a^{-1} M_p a) =
\frac{2\pi i}{k}\, d P\,{\rm tr}\, (dMM^{-1})^2 \equiv 0\,.
\lb{drhoex}
\ee
(Remember that the diagonal monodromy $M_p\,$ enters in $\O_q^{ex}\,$ through
\be
d M_p M_p^{-1} =
\frac{2\pi i}{k} \sum_{s=1}^n  d p_s e^s_s \equiv
\frac{2\pi i}{k}\,(dP\, \id + d \hat p )\,,
\lb{Mpex}
\ee
cf. (\ref{tr}), and ${\rm det}\, a\,$ is {\em not} set equal to
${\cal D}_q (p)\,.$)
Since $\r\,$ is {\em defined} by Eq. (\ref{rho-nN}),
it can be left unchanged in the extended phase space.
This is certainly true for the expression
(\ref{rhoinform})-(\ref{Ka}) provided we take -- as we will --
the {\em standard solution} of the modified CYBE (\ref{MCYBE})
\be
r_{12} = \sum_{\a >0} (e_\a\otimes e_{-\a} - e_{-\a}\otimes e_\a )\equiv
\sum_{\a >0} (e_{1\a} e_2^\a - e_1^\a e_{2\a} )\quad (e^\a := e_{-\a})
\lb{rstandard}
\ee
for which
\be
r\, e_{\pm\a} = \pm e_{\pm \a}\quad{\rm for}\quad \a >0\,,
\quad r\, h_i = 0 = r \id\,.
\lb{standaction}
\ee

\subsection{Poisson bivector for $\O_q^{ex}$}

We shall first establish the relation (\ref{chi})
which, according to (\ref{constr-ap}),
is sufficient to prove that the PB of $a^j_\a\,$
can be computed using the form
$\O_q^{ex}\,$ (\ref{Oqex}), (\ref{oqex}) on ${\cal M}_q^{ex}(n)\,.$
Eq. (\ref{chi}) follows from (\ref{oq})-(\ref{oqex}) and (\ref{drhoex})
(together with the subsequent argument), which imply
\be
\frac{\hat{\p}}{\p P} \r (a^{-1} M_p a) = 0\,.
\lb{ddPrho}
\ee

Similarly, one can deduce
\be
\sum_{s=1}^n {\hat V}^s_s\, \r (a^{-1} M_p a ) = 0\qquad\Rightarrow\qquad
-\frac{1}{n} \sum_{s=1}^n {\hat V}^s_s \, \O_q^{ex}= d P\,,
\lb{Vssonrho}
\ee
which extends the second equation (\ref{momentummaps}) to $q\ne 1\,.$

Using (\ref{rhoinform})-(\ref{Ka}), we can write
the extended symplectic form (\ref{Oqex}) as
\ba
&&\O_q^{ex} = \sum_{s=1}^n d p_s \Theta^s_s + \frac{k}{2\pi} \sum_{j\ne\ell , r\ne s}
\Theta^j_\ell \Theta^r_s [ (\o - X)^{-1} ]^{\ell s}_{j r} +\nonumber\\
&&+ \sum_{j , r\ne s, t\ne q}\, d p_j\,
\Theta^r_s \,X^{jt}_{jq}\, [(\o - X)^{-1} ]^{qs}_{tr} - \\
&&- \frac{\pi}{2k} \sum_{j\ne\ell} d p_j d p_\ell\,\left(\o^{jl} +
X^{j\ell}_{j\ell} +
\sum_{s\ne t, s'\ne t'}X^{js}_{jt} [ (\o - X)^{-1} ]^{tt'}_{ss'}
X^{s'\ell}_{t'\ell}\right)\nonumber
\lb{Oqex2}
\ea
where
\be
\o^{j\ell}:= i\,{\rm cotg} \frac{\pi}{k} p_{j\ell}\,,\quad \o^{nj}_{m\ell}
= -\o^{j\ell} \d^j_m \d^n_\ell = \o^{nj} \d^n_\ell \d^j_m
\lb{defs}
\ee
so that
\be
\o_q^{ex}(p) = \frac{\pi}{k} \sum_{j\ne\ell} \o^{j\ell} d p_j d p_\ell\,,
\quad \frac{A_+}{A_-} e^j_\ell =
- \o^{j\ell} e^j_\ell\equiv \o^{nj}_{m\ell} e^m_n
\lb{omegas}
\ee
(because $Ad_{M_p} e^j_\ell = q^{2p_{j\ell}} e^j_\ell$ )
and $X^{nj}_{m\ell}\,$ is defined as
\be
r^a e^j_\ell = - X^{nj}_{m\ell} e^m_n
\quad\Rightarrow\quad X_{12} = - Ad_{a_1 a_2} r_{12}\quad
(\, X_{12}=-X_{21}\, )\,.
\lb{defX}
\ee

To derive (\ref{Oqex2}), one uses (\ref{Onnq}), (\ref{rhoinform}),
(\ref{daa-1*}) and (\ref{Mpex}) as well as (\ref{defs}),
(\ref{omegas}). This allows to present $\O_q^{ex}\,$ as
\ba
&&\O_q^{ex}= - \o_q^{ex}(p) - \frac{k}{8\pi}\, {\rm tr} \,
\{ ( A_- da a^{-1}) (A_+ - K^a A_- ) (daa^{-1}) + \nonumber\\
&&+ 2\, dM_p M_p^{-1} (2 - K^a A_-)(daa^{-1}) +
d M_p M_p^{-1} K^a (d M_p M_p^{-1})\}
\lb{Oq}
\ea
Note that the only nonzero contribution of the diagonal
elements of $da a^{-1}\,$ comes through the term
\be
-\frac{k}{2\pi}\,{\rm tr}\, (d M_p M_p^{-1} da a^{-1}) =
\sum_{s=1}^n d p_s \Theta^s_s\,.
\lb{nonzero-diag}
\ee
One also uses relation (\ref{Ka}) and its corollary
\be
(A_+ - K^a A_- )r^a= K^aA_+-A_-\ \ \Rightarrow\ \
A_+ - K^a A_-  = 4 Ad_{M_p}\, (r^a A_- + A_+ )^{-1}
\lb{eqn}
\ee
as well as
\ba
&&\frac{k}{4\pi}\,{\rm tr}\, d M_p M_p^{-1} K^a A_- (d a a^{-1} ) =
\frac{k}{2\pi}\,{\rm tr}\,
d M_p M_p^{-1} r^a (r^a + \frac{A_+}{A_-} )^{-1} d a a^{-1} =\nonumber\\
&&= \sum_{j , r\ne s, t\ne q}\, d p_j\, \Theta^r_s \,
X^{jt}_{jq}\, [(\o - X)^{-1} ]^{qs}_{tr}
\lb{eqn2}
\ea
and
\ba
&&\ {\rm tr}\, d M_p M_p^{-1} K^a (d M_p M_p^{-1}) = {\rm tr}\,
d M_p M_p^{-1} r^a (d M_p M_p^{-1})+\nonumber\\
&&+ {\rm tr}\,
d M_p M_p^{-1} r^a [(1+\frac{A_-}{A_+} r^a)^{-1}-1] (d M_p M_p^{-1})\,,
\ea
where the second term in the right hand side gives
\ba
&&{\rm tr}\, (r^a d M_p M_p^{-1}) [(1+\frac{A_-}{A_+} r^a)^{-1}\,
\frac{A_-}{A_+} r^a ]
(d M_p M_p^{-1}) =\nonumber\\
&&=\frac{4\pi^2}{k^2} \sum_{j\ne\ell} d p_j d p_\ell\,
\sum_{s\ne t, s'\ne t'}X^{js}_{jt} [ (\o - X)^{-1} ]^{tt'}_{ss'}
X^{s'\ell}_{t'\ell}\,.
\ea

The PB derived from $\O_q^{ex}\,$ can be compactly written in terms
of the Poisson bivector
\ba
\lb{Poiss_k}
&&{\cal P} = \sum_{m=1}^n V^m_m\wedge\frac{\p}{\p p_m} +\\
&&+ \frac{\pi}{2k}
\left( \sum_{n\ne m} \o^{nm} (V^n_m\wedge V^m_n - V^m_m\wedge V^n_n ) -
\sum_{n,m,s,t} X^{ns}_{mt} V^m_n\wedge V^t_s \right)\nonumber
\ea
obeying the operator equation
\be
{\cal P}_{12} \, (\O_q^{ex})_{23} \, = \, I_{13}\,.
\ee
Here $I\,$ is the mixed $(1,1)$-tensor
\be
I = \sum_{j=1}^n (\frac{\p}{\p p_j}\otimes d p_j + V^j_j\otimes\Theta^j_j
) +
\sum_{j\ne\ell} V^j_\ell\otimes \Theta^\ell_j\,,
\ee
which plays the role of the identity operator in the space of $1$-forms
$\Theta\,,$ resp. vector fields $X\,$ in the sense
\ba
&&\Theta\, I := \sum_{j=1}^n \Theta (\frac{\p}{\p p_j})\,d p_j +
\sum_{j, \ell = 1}^n \Theta ( V^j_\ell )\, \Theta^\ell_j = \Theta\quad
(\,\Theta (X) \equiv {\hat X} \Theta\, )\,,\nonumber\\
&&I \, X := \sum_{j=1}^n \frac{\p}{\p p_j}\, d p_j (X) +
\sum_{j, \ell = 1}^n V^j_\ell \,\Theta^\ell_j (X) = X\,.
\lb{newfla}
\ea

We find, in particular,
\be
\{ a_1 , a_2 \} \equiv {\cal P}_{12} (a_1, a_2 )
= r_{12}(p) a_1 a_2 - \frac{\pi}{k} a_1 a_2 r_{12}
\lb{PBa}
\ee
where
\be
\lb{rp}
r(p)^{j\ell}_{j'{\ell}'}=
\left\{
\begin{array}{ll}
\, i \frac{\pi}{k}\, {\rm cotg} (\frac{\pi}{k} p_{j\ell})\, (\d^j_{j'}
\d^\ell_{{\ell}'} -
\d^j_{{\ell}'} \d^\ell_{j'} )\ & {\rm for}\ j\ne\ell\\
\, 0& {\rm for}\ j=\ell
\end{array}
\right.
\ee
and
\be
r^{\a\b}_{{\a}'{\b}'} = - \epsilon_{\a\b} \d^\a_{{\b}'} \d^\b_{{\a}'}
\lb{r}
\ee
(cf. (\ref{defX}) for the standard solution
(\ref{rstandard}), (\ref{standaction})).

The other two basic PB coincide with those in (\ref{PB1-1}) (and the Dirac
bracket $\{ a^j_a , p_\ell \}\,$ -- with (\ref{apD})).

The operators in the triple tensor product
${\C}^n\times{\C}^n\times{\C}^n\,$
\be
r^\pm_{ab} (p) = r_{ab} (p) \pm \frac{\pi}{k} C_{ab}\,,\qquad a,b=1,2,3\,
\ \ a< b
\lb{Cdynr}
\ee
satisfy the dynamical CYBE \cite{EV}
\be
[r^\pm_{12} (p), r^\pm_{13} (p) + r^\pm_{23} (p)]
+ [r^\pm_{13} (p), r^\pm_{23} (p) ] +
{\rm Alt}\, (d r^\pm ) = 0\,,
\lb{dynCYBE}
\ee
where
\be
{\rm Alt}\, (d r^\pm )
:= -i \sum_{j=1}^n \frac{\p}{\p p_j} \left(\,
{e^j_j}_1 r^\pm_{23} (p) - {e^j_j}_2 r^\pm_{13} (p) +{e^j_j}_3 r^\pm_{12} (p)\,
\right) \equiv {\rm Alt}\, (d r ) \,.
\lb{Altdr}
\ee
As the verification of (\ref{dynCYBE}) requires some work, we sketch the
main steps in the Appendix.

\section{Quantization}

\setcounter{equation}{0}
\renewcommand\theequation{\thesection.\arabic{equation}}

\subsection{Quantum exchange relations and their quasiclassical limit}

The exchange relations for the quantum matrix algebra
-- which we shall again denote by ${\cal M}_q\,$ --
have been derived
earlier on the basis of an analysis of the braiding properties of
$SU(n)_k\,$ WZNW $4$-point blocks \cite{FHIOPT, HST} satisfying the
Knizhnik-Zamolodchikov equations \cite{KZ, TK}. They have the form
\cite{FHIOPT}
\ba
\lb{QMA1}
&&[ q^{p_{ij}} , q^{p_{k\ell}} ] = 0\,,\quad
q^{p_{ij}} a^\ell_\a = a^\ell_\a q^{p_{ij} +\d^\ell_i -\d^\ell_j}\,,\\
&&{\hat R}(p)^{\pm 1} a_1 a_2 = a_1 a_2 {\hat R}^{\pm 1}\,,
\lb{QMA2}
\ea
where
\ba
\lb{hatR}
&&(q^{\frac{1}{n}} {\hat R})^{\pm 1}_{i\, i+1} =
q^{\pm 1} \id_{i\, i+1} - A_{i\, i+1}\,,\quad
A^{\a_1\a_2}_{\b_1\b_2} = q^{\epsilon_{\a_2\a_1}}
\d^{\a_1}_{\b_1} \d^{\a_2}_{\b_2} - \d^{\a_1}_{\b_2} \d^{\a_2}_{\b_1}\,,\\
&&q^{\epsilon_{\a_2\a_1}} =
\left\{
\begin{array}{ll}
\, q^{-1} \ & {\rm for}\ \a_1 < \a_2\\
\, 1& {\rm for}\ \a_1 = \a_2\\
\, q& {\rm for}\ \a_1 > \a_2
\end{array}
\right.
\,,\quad q = e^{-i\frac{\pi}{h}}\,,\ \ h=k+n
\lb{qepsilon}\\
&&(q^{\frac{1}{n}} {\hat R}(p))^{\pm 1}_{i\, i+1} =
q^{\pm 1} \id_{i\, i+1} - A_{i\, i+1} (p)\,,\\
&&A^{i_1i_2}_{j_1j_2} (p) = \frac{[p_{i_1 i_2}-1]}{[p_{i_1i_2}]}
(\d^{i_1}_{j_1}\d^{i_2}_{j_2} - \d^{i_1}_{j_2}\d^{i_2}_{j_1} )\,.\nonumber
\lb{hatRp}
\ea
Both $A_{i\, i+1} =: A_i\,$ and $A_{i\, i+1} (p) =: A_i (p)\,$
satisfy the Hecke algebra relations
\be
A_i A_{i+1} A_i - A_i =
A_{i+1} A_i A_{i+1} - A_{i+1}\,,\ \, A_i^2 = [2] A_i\,,\ \ [ A_i , A_j ] =
0 \ \ {\rm for}\ \ |i-j|>1\,.
\lb{HeckeA}
\ee
It remains to verify that the quasiclassical limit of these relations
indeed reproduces the PB relations of Section 3.

One can introduce two deformation parameters: $\frac{1}{k}\,$ and the
(implicit in common notation) Planck constant $\hbar\,$ -- see \cite{AF}. If
one ascribes to the physical quantities ${\tilde k}\,$ and ${\tilde p}\,$
the dimension of action, then our dimensionless numbers $k\,$ and $p\,$
shall be written as $k = \frac{\tilde k}{\hbar}\,$ and $p = \frac{\tilde
p}{\hbar}\,.$ We shall distinguish the quasiclassical limit ($\hbar\to
0\,$) from the undeformed limit ($k\to\infty\,$) without using the
parameter $\hbar\,,$ by characterizing the second one by
\be
k \to \infty\,,
\quad p_{j\ell}\ \ {\rm finite}\,,
\quad \frac{p_{j\ell}}{k}\to 0\,,
\lb{kinftylim}
\ee
while setting for the first one of interest
\be
\frac{k}{n} \to \infty\,,
\quad p_{j\ell}\to \infty\,,
\quad\frac{p_{j\ell}}{k}\ \ {\rm finite}\quad (\, j<\ell\, )\,.
\lb{quasicl}
\ee
The substitution of the {\em level} $k\,$ by the {\em height} $h=k+n\,$ in
the quantum expression for $q\,$ (\ref{qepsilon}) is consistent with
(\ref{quasicl}) but we are only aware of an explanation of its necessity
that uses the full (with infinite number of degrees of freedom)  WZNW
model which involves the Sugawara formula expressing the stress energy
tensor as a norm square of the $SU(n)\,$ current (see \cite{KZ, T}).

Let $P\,$ be the permutation operator for either set of indices,
$j,\ell,\dots\,$ or $\a ,\b ,\dots\,:$
\be
P_{12} = \left( P^{j_1 j_2}_{\ell_1 \ell_2} \right) = \left( \d^{j_1}_{\ell_2}\d^{j_2}_{\ell_1} \right)\quad{\rm or}\quad
P_{12} = \left( P^{\a_1 \a_2}_{\b_1 \b_2} \right) = \left( \d^{\a_1}_{\b_2} \d^{\a_2}_{\b_1} \right)
\lb{P}
\ee
and let $\id_{12}\,$ be the corresponding
unit operator (e.g., $\id^{j_1 j_2}_{\ell_1\ell_2} =
\d^{j_1}_{\ell_1}\d^{j_2}_{\ell_2}\,$).
Then we can write
\ba
\lb{R}
&&R^{\a_1\a_2}_{\b_1\b_2} = ({\hat R} P)^{\a_1\a_2}_{\b_1\b_2} =
\bq^{\frac{1}{n}}\left(
(q-q^{\epsilon_{\a_2\a_1}} ) P^{\a_1\a_2}_{\b_1\b_2}
+ \id^{\a_1\a_2}_{\b_1\b_2} \right)\,,\\
&&R(p)^{j_1 j_2}_{\ell_1\ell_2} = ({\hat R}P)^{j_1 j_2}_{\ell_1\ell_2} =
\bq^{\frac{1}{n}}
\left( \frac{q^{p_{j_1 j_2}}}{[p_{j_1 j_2}]}
P^{j_1 j_2}_{\ell_1\ell_2} +
\frac{[p_{j_1 j_2} -1]}{[p_{j_1 j_2}]} \id^{j_1 j_2}_{\ell_1\ell_2}\right)\,.
\lb{Rp}
\ea
Setting now
\ba
&&q=1-i\frac{\pi}{k} + {\cal O}(\frac{\pi^2}{k^2})\qquad \
(\ \bq^{\frac{1}{n}} = 1+ i\frac{\pi}{nk} +
{\cal O}(\frac{\pi^2}{k^2})\ ) \nonumber\\
&&\frac{[p-1]}{[p]} = 1-\frac{\pi}{k} {\rm cotg}\,(\frac{\pi}{k} p) +
{\cal
O}(\frac{\pi^2}{k^2})\,,
\lb{expansion}
\ea
we find
\be
R_{12} = \id_{12} + i\frac{\pi}{k} r^-_{12}
+{\cal O}(\frac{\pi^2}{k^2}\,)\,,\ \
R(p)_{12} = \id_{12} + i\, r^-_{12} (p)
+{\cal O}(\frac{\pi^2}{k^2}\,)\,,
\lb{expanR}
\ee
where
\ba
\lb{r-}
r^-_{12} = r_{12} - C_{12}\,,\quad
r^{\a_1\a_2}_{\b_1\b_2} = -\epsilon_{\a_1\a_2}
P^{\a_1\a_2}_{\b_1\b_2}\,,\quad C_{12} = P_{12} - \frac{1}{n} \id_{12}\,,\\
r^-_{12}(p) = r_{12} - C_{12}\,,\quad
r(p)^{j_1 j_2}_{\ell_1\ell_2} = i\frac{\pi}{k} {\rm cotg}\,
(\frac{\pi}{k} p_{j_1 j_2})
(\d^{j_1}_{\ell_1} \d^{j_2}_{\ell_2} - \d^{j_1}_{\ell_2} \d^{j_2}_{\ell_1} )\,.
\lb{rp-}
\ea
The reason why we are keeping the factor $\frac{\pi}{k}\,$ in
the definition of $r_{12}(p)\,$ is that it has
a nonzero undeformed limit since
\be
\lim_{k\to\infty}\,\frac{\pi}{k}\,{\rm cotg}\, (\frac{\pi}{k} p)
= \frac{1}{p}\,.
\lb{limes}
\ee

Taking into account that $[C_{12} , a_1 a_2 ] = 0\,,$
we thus recover the PB relations of Section 3.2.

\subsection{$U_q(s\ell_n )\,$ symmetry of the exchange relations}

Let, for $G_0\ni M \equiv (M^i_j )_{i,j = 1}^n\,,\ \ M_n^n\ne 0 \ne {\rm
det}\,\left(\matrix{M_{n-1}^{n-1}&M^{n-1}_n\cr
M^n_{n-1}& M_n^n} \right)\,$ etc. and
\ba
M = q^{\frac{1}{n}-1} M_+ M_-^{-1}\,,\
M_+ = N_+ D\,,\  M_-^{-1} = N_- D\,,\quad D= (d_\a \d^\a_\b )\,,
\lb{M+-}\\
N_+ = \left(
\matrix{1&f_1&f_{12}&\dots\cr 0&1& f_2&\dots\cr 0&0&1&\dots\cr
\dots&\dots&\dots&\dots}\right) ,\
N_- = \left(
\matrix{1&0&0&\dots\cr e_1&1& 0&\dots\cr e_{21}&e_2&1&\dots\cr
\dots&\dots&\dots&\dots}\right) ,
\lb{N+-}
\ea
where the common diagonal matrix $D\,$ has unit determinant: $ d_1 d_2
\dots
d_n = 1\,.$ It can be deduced from
(\ref{QMA1}), (\ref{QMA2}) and $M = a^{-1} M_p a\,$ that
\be
[ {\hat R}^{\pm} , M_{2\pm} M_{1\pm} ] = 0\,,\quad
{\hat R} M_{2 -} M_{1 +}  = M_{2 +} M_{1 -}  {\hat R} \,.
\lb{RM+-}
\ee
It is known that Eqs. (\ref{RM+-}) for the matrices $M_\pm\,$ are
equivalent to the defining relations of the quantum universal enveloping
algebra $U_q := U_q (s\ell_n )\,$ \cite{CP}
that is paired by duality to $Fun\, (SL_q(n))\,$ \cite{FRT}.
The Chevalley generators of $U_q \,$
are related to the elements of the matrices (\ref{M+-}), (\ref{N+-}) by
(\cite{FRT}, see also \cite{FHT3})
\ba
&&d_i = q^{\L_{i-1}-\L_i}\qquad (i=1,\dots , n\,,\ \L_0 = 0 = \L_n
)\,,
\nonumber\\
&&e_i = (\bq - q) E_i\,,\qquad f_i = (\bq -q ) F_i\,,\nonumber\\
&&(\bq -q) f_{12} = f_2 f_1 - q f_1 f_2 = (\bq -q)^2 (F_2 F_1 - q F_1 F_2
)\ \ {\rm etc.},\nonumber\\
&&(\bq -q) e_{21} = e_1 e_2 - q e_2 e_1 = (\bq -q)^2 (E_1 E_2 - q E_2 E_1
)\ \ {\rm etc.}
\lb{Uq}
\ea
Here $\L_i\,$ are the fundamental co-weights of $s\ell (n)\,$ related to
the co-roots $H_i\,$ by $H_i = 2 \L_i - \L_{i-1} - \L_{i+1}\,;\ E_i\,$ and
$F_i\,$ are the raising and lowering operators satisfying
\ba
&&[E_i , F_j ] = [H_i ] \d_{ij}\,,\ \
q^{\L_i} E_j = E_j q^{\L_i+\d_{ij}}\,,\
q^{\L_i} F_j = F_j q^{\L_i-\d_{ij}}\,,\nonumber\\
&&[ E_i , E_j ] = 0 = [ F_i , F_j ]\qquad{\rm for}\quad |j-i|\ge
2\,,\nonumber\\
&&[2]X_i X_{i\pm 1} X_i = X_{i\pm 1} X_i^2 + X_i^2 X_{i\pm 1}\qquad{\rm
for}\quad X=E,F\,.
\lb{Uqn}
\ea
The exchange relations (\ref{QMA1}), (\ref{QMA2}) imply
\be
M_{1\pm} P a_1 = a_2 {\hat R}^{\mp 1} M_{2\pm}
\lb{M+-a}
\ee
(see \cite{FHIOPT}). It follows that these exchange relations are
invariant under the coaction of $U_q \,,$
\ba
&&[ E_a , a^i_\a ] = \d_{a\,\a-1} a^i_{\a -1} q^{H_a}\,,\quad
[ q^{H_a} F_a , a^i_\a ] = \d_{a\,\a} q^{H_a} a^i_{\a+1}\,,
\nonumber\\
&&q^{H_a } a^i_\a = a^i_\a q^{H_a + \d_{a\,\a}-\d_{a\,\a-1}}\,,\quad
{\footnotesize a} = 1,\dots ,n-1\,.
\lb{coaction}
\ea
We note that the centralizer of $q^{p_i}\ \ ( \prod_{i=1}^n q^{p_i} = 1
)\,$ in the algebra (\ref{QMA1}), (\ref{QMA2}) (i.e., the maximal
subalgebra commuting with all $q^{p_i}\,$) is spanned by $U_q \,$
over the field ${\Bbb Q} (q , q^{p_i})\,$ of rational functions of
$q^{p_i}\,.$

\subsection{Operator realization}

We shall sketch and briefly discuss the finite-dimensional Fock-like space
realization of the quantum matrix algebra of \cite{FHIOPT}.

The "Fock space" ${\cal F}\,$ and its dual ${\cal F}'\,$
are defined as ${\cal M}_q$-modules with $1$-dimensional $U_q$-invariant
subspaces of multiples of (non-zero) bra and ket vacuum vectors
${\cal h} 0 |\,$ and $ | 0 {\cal i}\,$ (such that
${\cal h} 0 | {\cal M}_q = {\cal F}'\,,\ {\cal M}_q | 0 {\cal i} =
{\cal F}\,$) satisfying
\ba
&&a^i_\a | 0 {\cal i} = 0\quad{\rm for}\quad i>1\,,\quad\quad {\cal h} 0 |
a^j_\a = 0 \quad{\rm for}\quad j<n\,,\nonumber\\
&&q^{p_{ij}} | 0 {\cal i} = q^{j-i} | 0 {\cal i}\,,\quad\quad
{\cal h} 0 | q^{p_{ij}} = q^{j-i} {\cal h} 0 |\,,\nonumber\\
&&(X - \e (X)) | 0 {\cal i} = 0 = {\cal h} 0 | ( X - \e (X))\,,
\quad\forall \,X \in U_q\,,
\lb{Fock}
\ea
with $\e (X)\,$ the counit. The duality between
${\cal F}\,$ and ${\cal F}'\,$ is established by a bilinear pairing
${\cal h}\, .\, |\, .\, {\cal i}\,$ such that
\be
{\cal h} 0 | 0 {\cal i} = 1\,, \quad
{\cal h} \Phi | A | {\Psi} {\cal i}
= {\cal h} \Psi | A' | {\Phi}  {\cal i}
\lb{transp}
\ee
where $A\ \to \ A'\,$ is a linear antiinvolution ({\em transposition})
of ${\cal M}_q\,$ defined for generic $q\,$ by
\be
{\cal D}_i(p) (a^i_\a )' = {\tilde a}^\a_i :=\frac{1}{[n-1]!}
{\cal E}^{\a\a_1\dots\a_{n-1}} \e_{i i_1\dots i_{n-1}}
a^{i_1}_{\a_1}
\dots a^{i_{n-1}}_{\a_{n-1}}\,,\ \ \ (q^{p_i} )' = q^{p_i}\,.
\lb{deftransp}
\ee
Here ${\cal D}_i (p)\,$ stands for the product
\be
{\cal D}_i(p) = \prod_{j<\ell ,\, j\ne i\ne\ell} [ p_{j\ell} ]\qquad
(\,\Rightarrow\ [ {\cal D}_i (p) , a^i_\a ] = 0 = [ {\cal D}_i (p) ,
{\tilde a}^\a_i ]\, )\,;
\lb{Dip}
\ee
for the definition of the $U_q$- and,
respectively, the "dynamical" Levi-Civita tensors
${\cal E}^{\a_1\a_2\dots\a_{n}}\,,\  \e_{i_1 i_2\dots i_{n}}\,$
see \cite{HIOPT, FHIOPT}.
The antiinvolution (\ref{deftransp})
extends the known transposition of $U_q\,$
determined by its action on the Chevalley generators
(see Section 3 of \cite{FHT3}),
\be
{E_i}' = F_i\, q^{H_i -1}\,,\quad
{F_i}' = q^{1-H_i} E_i\,,\quad (q^{H_i} )' = q^{H_i}\,,
\lb{transpChev}
\ee
to the quantum matrix algebra (cf. Section 3.1 and Appendix B of
\cite{FHIOPT}).

The space ${\cal F}\,$ admits a canonical basis of weight vectors whose
inner product can be computed (see Section 3.2 of \cite{FHIOPT}). For
$n=2\,$ the basis has the simple form
\be
| p , m {\cal i} = (a^1_1 )^m (a^1_2 )^{p-1-m} \ | 0 {\cal i}\,,\qquad
0\le m\le p-1\quad (p\equiv p_{12} )\,,
\lb{bas2}
\ee
and the inner product is given by
\be
{\cal h} p' , m' | p , m {\cal i} = \d_{p p'} \d_{m m'} \bq^{m(p-1-m)}
[m]! [p-1-m]!\,.
\lb{inner2}
\ee

For the deformation parameter $q\,$ appearing in (\ref{qepsilon}),
\be
q = e^{-i\frac{\pi}{h}}\,,\ \ h=k+n\qquad \Rightarrow\qquad q^h = -1
\lb{q-quantum}
\ee
i.e., $q\,$ a (here, even) root of unity, the Fock space has an infinite
dimensional $U_q\,$ invariant subspace of {\em null vectors} orthogonal to
any vector in ${\cal F}\,.$ In the $n=2\,$ case all null vectors belong to
the set ${\cal I}_h\, | 0 {\cal i}\,$
%% or ${\cal h} 0 | {\cal I}_h\,$
where ${\cal I}_h\,$ is the ideal generated by
$[ h p ]\,,\ [ h H ]\,,\ q^{hp}+q^{hH}\,,\ ( a^i_\a )^h\,, i,\a=1,2\,.$
%% and by the $h$-th powers of the $a^i_\a\,$ (or, equivalently,
%% by the $h$-th powers of the ${\tilde a}_i^\a\,$).
The definition of the
ideal ${\cal I}_h\,$ can be generalized to any $n\ge 2\,$ assuming that it
includes the $h$-th powers of all minors of the quantum matrix $( a^i_\a
)\,.$ For $n=2\,$ the factor space ${\cal F}_h\,$ is spanned by vectors of
the form (\ref{bas2}) with $0 < p < 2h\,$ and $m\,$ in the range $0\le m
\le p-1\,,$ for $1\le p \le h\,,$ and $p-h \le m \le h-1\,$ for $h+1\le p
\le 2h-1\,.$ It splits into a direct sum of $2h-1\,$ irreducible
representations of $U_q (s\ell_2 )\,$ of total dimension $h^2\,.$

For general $n\,$ and generic $q\,$ ($q\,$ not a root of unity)  the space
${\cal F}\,$ has been proven to be a model space for $U_q\,$ (see Section
3.1 of \cite{FHIOPT}). The question of what should be viewed as a model
space for the {\em reduced} $U_q\,$ ($U_q\,$ factored by its maximal ideal)
for $q\,$ satisfying (\ref{q-quantum}) appears to be unsettled. If we
define it as the direct sum of {\em integrable representations}
(those with $0< p < h\,,$ for $n=2\,$) of multiplicity $1\,,$
then the question arises whether there is a natural (say, a BRS type)
procedure that would reduce ${\cal F}_h\,$ to such a sum.
A BRS procedure was introduced in \cite{DVT1, DVT2}
for the tensor product of two copies of ${\cal F}_h\,$ -- corresponding to
the left and right movers' zero modes of a $SU(2)\,$ WZNW model -- but
this changes the problem.

\section{Concluding remarks}

%% The systematic study of the chiral zero modes' phase space of the 
%% SU(n) WZNW model reveals some unexpected features.

We have tried to make the present study of the chiral zero modes' 
phase space reasonably self-contained and have, hence, included 
some known material. It may be, therefore, useful to list at 
this point what appears to us as the main new features in our treatment.

We find explicitly the correspondence between the WZ term 
$\rho (a^{-1} M_p a)\,$ (rendering the zero modes' symplectic form
(\ref{Onnq}) closed) and the solutions of the CYBE.

It is essential for the present treatment of the $SU(n)\,$ case
that the determinant ${\rm det}\, ( a^i_\a )\,$ of the zero modes'
$n\times n\,$ matrix is set equal to 
%% 2. would propose to replace "a specific function" by: 
an (essentially unique) pseudoinvariant $q$-polynomial 
in the $su(n)\,$ weights - see (\ref{D}).
%% a specific function, see (\ref{D}), of the
%% "momenta" $\{ p_i \}_{i=1}^n\,$ (the latter defining a vector in
%% a Weyl alcove of the $su(n)\,$ weight space).
%% whose quantum eigenvalues give the dominant weights of SU(n).
Accordingly, the symplectic form (\ref{Onnq}) in the $(n-1)(n+2)$-dimensional 
zero modes' phase manifold ${\cal M}_q\,$ necessarily contains, for $n>2\,,$ 
a term ${\omega}_q (M_p)\,$ depending only on the diagonal monodromy.

The counterpart of ${\omega}_q (M_p)\,$ in the symplectic form of
the chiral WZNW model with diagonal monodromy, being closed by itself, 
is often omitted. This additional term is necessary in order 
to reproduce upon quantization the basic exchange relations involving 
the dynamical $R$-matrix of \cite{GN, AF, BF, FHIOPT}.
%%[36] [1] [14] [31].  

The expression for $\omega_q\,$ is simpler -- and easier to derive --
in the extended ($n(n+1)$-dimensional) phase space ${\cal M}_q^{ex}\,$
spanned by $p_i\,$ and $a^j_\a\,,\ i,j,\a = 1,\dots ,n\,,$
%% /see expression above Eq. (2.21) of our hep-th/0211154/, 
the form $\omega_q^{ex}\,$ (\ref{oqex})
%% (2.28)
being nontrivial even for $n=2\,$ (yielding, in the undeformed limit, the 
standard symplectic structure on ${\C}^2\,$ viewed as a K\"ahler manifold in that case). 

The Dirac brackets of the physically interesting quantities $a^j_{\alpha}\,$
and $p_{j\ell}\,$ coincide with their Poisson brackets since they Poisson commute with 
one of the constraints and can be, hence, derived working in the (more symmetric)
extended phase space.

The expression (\ref{Poiss_k})
%% (3.30) 
for the Poisson bivector in ${\cal M}_q^{ex}\,$ allows to directly 
compute the Poisson brackets of interest.

The quantum theory of chiral zero modes has been only briefly 
reviewed in Section 4
concluding with the formulation of an open problem related 
to the concept of a 
model space for the quantum universal enveloping algebra
$U_q(s\ell_n )\,$ for $q\,$ a root of unity.

\section*{Acknowledgements}

P.F. acknowledges the support of the Italian Ministry of Education,
University and Research (MIUR). L.H. thanks Dipartimento di Fisica Teorica
(DFT) dell' Universit\`a di Trieste and Istituto Nazionale di Fisica
Nucleare (INFN), Sezione di Trieste; both he and I.T. also acknowledge the
hospitality and support of the Erwin Schr\"odinger International Institute
for Mathematical Physics (ESI), Vienna. This work is supported in part by
the Bulgarian National Council for Scientific Research under contract
F-828.

\section*{Appendix A}

\setcounter{equation}{0}
\renewcommand\theequation{A.\arabic{equation}}

We begin by reproducing the properties of the polarized Casimir operators
$C_{mn}\,$ relevant for the proof of the CYBE (for both constant and
"dynamical" -- i.e., $p$-dependent -- $r^\pm\,$):
\be
[ C_{12}, C_{13} + C_{23} ] = 0 = [ C_{12} + C_{13} , C_{23} ]\,.
\lb{CCC}
\ee
For $G=SU(n)\ \ C_{12}\,$ is given, essentially, by the permutation
operator (\ref{P}):
\be
C_{12} = P_{12} - \frac{1}{n} {\id}_{12}\,,\ \ {\rm or}\ \
C^{j_1 j_2}_{\ell_1 \ell_2}
= \d^{j_1 j_2}_{\ell_2 \ell_1} - \frac{1}{n} \d^{j_1 j_2}_{\ell_1 \ell_2}
\qquad (\, \d^{jk}_{\ell m} := \d^j_\ell \d^k_m \, )\,,
\lb{C12}
\ee
and Eq. (\ref{CCC}) gives
\be
[ C_{12} , C_{13} + C_{23} ] + [ C_{13} , C_{23} ] = - [ C_{12} , C_{23} ]
= \left(\,
\d^{j_1 j_2 j_3}_{\ell_2\ell_3\ell_1}
- \d^{j_1 j_2 j_3}_{\ell_3\ell_1\ell_2}
\,\right)\,.
\lb{CCCeq}
\ee
Next we verify that the mixed ($r$-$C$) terms in the CYBE
(\ref{dynCYBE}) (or (\ref{sCYBE})) vanish,
\be
[ r_{12} (p) , C_{13} + C_{23} ] +
[ r_{12} (p) , C_{23} - C_{12} ] +
[ C_{12} + C_{13} , r_{23} (p) ] = 0\,,
\lb{[rC]}
\ee
using, e.g., the general identities $P_{ab} r_{bc} = r_{ac} P_{ab}\,$ for
$a,b,c\,$ all different, as well as skewsymmetry of $r_{ab}=-r_{ba}\,.$
Computing the sum of commutators in (\ref{dynCYBE}), we find
\ba
&&\left(\, [ r_{12}(p) , r_{13} (p) + r_{23} (p) ] + [ r_{13} (p) , r_{23}
(p) ] \,\right)^{j_1 j_2 j_3}_{\ell_1\ell_2\ell_3}
 =
\nonumber\\
&&= \frac{\pi^2}{k^2}
( ( c_{j_1 j_2} + c_{j_2 j_3} ) c_{j_1 j_3} - c_{j_1 j_2} c_{j_2 j_3} )
\left(\, \d^{j_1 j_2 j_3}_{\ell_2\ell_3\ell_1} -
\d^{j_1  j_2 j_3}_{\ell_3\ell_1\ell_2}\, \right)\,,
\lb{commut}
\ea
where
\be
c_{j\ell}:= {\rm cotg} \frac{\pi}{k} p_{j\ell} = - c_{\ell j}\,,\ \
j\ne\ell\,,\ \qquad c_{\ell\ell} := 0\,.
\lb{cjl}
\ee
On the other side, (\ref{Altdr}) gives
\be
{\rm Alt} (dr) =\frac{\pi}{k}
 ( \d_{j_1 j_2} c'_{j_2 j_3} + \d_{j_2 j_3} c'_{j_1 j_3}
+ \d_{j_1 j_3} c'_{j_1 j_2} )
\left(\, \d^{j_1 j_2 j_3}_{\ell_2\ell_3\ell_1} -
\d^{j_1  j_2 j_3}_{\ell_3\ell_1\ell_2}\, \right)\,,
\lb{Alt2}
\ee
where
\be
c'_{j\ell}:= - \frac{\pi}{k} \frac{1}{\sin^2 \frac{\pi}{k} p_{j\ell}} =
c'_{\ell j}\,,\ \
j\ne\ell\,,\ \qquad c'_{\ell\ell} := 0\,.
\lb{cjlprime}
\ee
To prove (\ref{dynCYBE}), it suffices to combine
(\ref{CCCeq})-(\ref{cjlprime}) with one of the following relations
(depending on whether all the three indices $j_1,j_2,j_3\,$ are different
or not):
\ba
&&
({\,\rm cotg}\,\a + {\rm cotg}\,\b\,
)\, {\rm cotg}\, (\a +\b ) - {\rm cotg}\,\a\,{\rm cotg}\,\b = - 1\,,
\lb{cotg1}\\
&&
{\rm cotg}^2 \,\a - \frac{1}{{\rm sin}^2\,\a} = - 1\,.
\lb{cotg2}
\ea
%%%%%%%%%%%%% BIBLIOGRAPHY - (~CMP style) %%%%%%%%%%%%%%%%%%%

\end{document}